\documentclass[%
 reprint,
amsmath,
amssymb,
aps, 
pra,
]{revtex4-2}

\usepackage{xcolor}
\usepackage{lipsum}
\usepackage{booktabs}
\usepackage{algpseudocode}
\usepackage{float}
\usepackage[normalem]{ulem}
\usepackage{amssymb,amsmath,amsfonts}
\usepackage{makecell}
\usepackage{graphicx} 
\usepackage{dcolumn} 
\usepackage{bm} 
\usepackage{hyperref} 

\hypersetup{
    colorlinks = true,
    citecolor = blue,
    linkcolor = blue,
    filecolor = magenta,      
    urlcolor = .,
    }

\begin{document}

\preprint{APS/123-QED}

\title{
Shadow measurements for feedback-based quantum optimization}

\author{Leticia Bertuzzi}
    \email{letyh.bz@gmail.com}
    \affiliation{Departamento de Física, Universidade Federal de Santa Catarina, 88040-900, Florianópolis, Brazil}

\author{João P. Engster}
    \email{engsterjp@gmail.com}
    \affiliation{Departamento de Física, Universidade Federal de Santa Catarina, 88040-900, Florianópolis, Brazil}

\author{Evandro C. R. da Rosa}
    \email{evandro.crr@posgrad.ufsc.br}
    \affiliation{Departamento de Informática e Estatística, Universidade Federal de Santa Catarina, 88040-900, Florianópolis, Brazil}    

\author{Eduardo I. Duzzioni}
    \email{eduardo.duzzioni@ufsc.br}
    \affiliation{Departamento de Física, Universidade Federal de Santa Catarina, 88040-900, Florianópolis, Brazil}

\date{\today}

\begin{abstract}
Improving the performance of quantum algorithms is a fundamental task to achieve quantum advantage. In many cases, extracting information from quantum systems poses an important challenge for practical implementations in real-world quantum computers, given the high resource cost of performing state tomography. In this scenario, randomized measurements emerged as a promising tool. In particular, the classical shadows protocol allows one to retrieve expected values of low-weight Pauli observables by performing only local measurements. In this paper, we present an implementation of the recently introduced Feedback-based algorithm for quantum optimization (FALQON) with the Ket quantum programming platform, for solving the MaxCut optimization problem. We employ classical shadows for the feedback routine of parameter estimation and compare this approach with the direct estimation of observables. Our results show that depending on the graph geometry for the MaxCut problem, the number of measurements required to estimate expected values of observables with classical shadows can be up to 16 times lower than with direct observable estimation. Furthermore, by analyzing complete graphs, we numerically confirm a logarithmic growth in the required number of measurements relative to the number of observables, reinforcing that classical shadows can be a useful tool for estimating low-locality Pauli observables in quantum algorithms. 
\end{abstract}

\maketitle


\section{Introduction}
\label{sec: introduction}
Quantum optimization algorithms aim to find the best solution to a problem from a set of possible solutions. Some examples include the quantum adiabatic evolution~\cite{farhi2000quantum,farhi2001quantum}, quantum least squares fitting~\cite{wiebe2012quantum}, quantum approximate optimization algorithm (QAOA)~\cite{farhi2014quantum}, and quantum semidefinite programming~\cite{Brandao2017}. The QAOA is a hybrid quantum-classical approach designed to solve combinatorial optimization problems. The evolution operator comprises two alternating unitary operators; one encodes the solution into the ground state of the target Hamiltonian, while the mixing operator explores the solution space. It iteratively optimizes circuit parameters using classical methods to approximate the solution. However, QAOA faces challenges such as barren plateaus~\cite{larocca2405review}, which emerge during the classical optimization process. This process by itself can be computationally expensive and may get stuck in local minima.

Recently proposed, the \textbf{F}eedback-based \textbf{AL}gorithm for \textbf{Q}uantum \textbf{O}ptimizatio\textbf{N} (FALQON) is purely quantum, \textit{i.e.}, it eliminates the need for classical parameter optimization~\cite{Magann2022}. Instead, it uses a feedback loop to adjust parameters in real-time iteratively. By leveraging a specific feedback rule, FALQON applies time-dependent controls to navigate toward the optimal solution. This approach simplifies the optimization process, reduces computational overhead compared to QAOA, and is well-suited for near-term quantum devices.

Beyond designing optimal feedback rules, one of the major concerns related to FALQON is the measurement overhead. Estimating observables at each step for feedback rules can demand a significant number of measurements, especially for larger problem instances. Despite recent publications regarding applications~\cite{rahman_feedback-based_2024-1, pexe_using_2024, larsen_feedback-based_2024, li_simulation_2024, scott_hybrid_2024} and improvements~\cite{Arai2025, Malla2024} of FALQON, the measurement process is often overlooked, with results simulated using the exact values of observables - which can never be obtained in practice, since it would require an infinite number of measurements. In this article, we simulate a scenario with finite resources, estimating the number of measurements necessary for the algorithm to converge.

An efficient method for obtaining the expected values of Pauli observables is the classical shadows technique~\cite{Huang2020}, which involves performing random measurements of local observables on the system's state. We employ the classical shadows technique as a replacement for a direct estimation procedure to obtain the expected values of the observables used to describe the Lyapunov function. As a case study, we apply this approach to solve the MaxCut problem for graphs with up to ten vertices. Our results highlight the superior efficiency of classical shadows over direct measurements, demonstrating a reduction of up to sixteen thousand measurements in the examples analyzed.

In Section \ref{sec: shadows} and \ref{subsec: falqon review} we review the classical shadows method and FALQON, respectively. The introduction of classical shadows measurements as a subroutine of FALQON is shown in Section \ref{subsec: mod falqon}. The results and discussion are presented in Section \ref{sec: results} and the final remarks in Section \ref{sec: conclusions}. 

\section{Classical shadows}
\label{sec: shadows}
Retrieving information from quantum systems is a central task in quantum computing. Usually, quantum algorithms encode the solution to a problem in the complex amplitudes of a quantum state. Such information is contained in the elements of the density operator $\rho$ describing the quantum state, so that to retrieve the output of the algorithm one needs to perform quantum state tomography (QST) - a procedure that reconstructs the elements $\rho_{ij}$ from measurements on the given state~\cite{Paris2004, James2001}. Without any prior knowledge of the state of the system, performing QST up to an additive error $\epsilon$ scales with $\mathcal{O}(4^n/\epsilon^2)$ \cite{Cardoso2021} for a system of $n$ qubits. In other words, the number of observables to be estimated scales exponentially with the size of the system.
However, performing a full QST is often unnecessary, as many applications require only accurate approximations of physical quantities rather than the complete state description. Such quantities are usually linear or quadratic functions of the density operator - for instance, expected values of observables and entanglement entropies, respectively~\cite{Huang2020}. Although subtle, this consideration was crucial in the development of the shadow tomography protocol~\cite{Aaronson2017}, resulting in a logarithmic sample complexity in the number of observables, albeit burdensome in terms of hardware (requiring exponentially long circuits acting collectively on the copies of $\rho$).

Inspired by the insights from shadow tomography, Huang \textit{et al.} proposed the protocol of classical shadows~\cite{Huang2020}, allowing one to efficiently estimate $L$ linear functions of $\rho$ from local measurements, with sample complexity only logarithmic in the number of estimated observables - see Eq. \eqref{eq: shadow sample complexity}. The classical shadows technique has proven to be of great value, much for its favorable scaling and versatility, given that it involves a randomized measurement procedure. The protocol has been applied in a wide range of problems, varying from optical experimental implementations~\cite{Struchalin2021, Zhang2021, Liu2022}, efficient model training in quantum machine learning~\cite{Jerbi2024, Abbas2023}, fermionic systems~\cite{Zhao2021}, entanglement verification~\cite{Elben2020, Vermersch2024}, and the estimation of Fischer information in quantum processors~\cite{Vitale2024, Rath2021}.

The protocol consists of constructing a classical representation of an unknown quantum state by performing projective measurements on randomly chosen bases. Such bases are specified by unitary operators that rotate the system from the computational $Z$-basis to the Pauli bases $X$, $Y$, or $Z$ (identity operator). This characterizes the so-called Pauli measurements, which are particularly suitable for estimating Pauli observables. For each measurement, one unitary is sampled from the Pauli group with a uniform probability distribution. Given the selected measurement basis and its respective measurement outcome, we can construct an estimator $\hat{\rho}$ of the density operator of the system. Then, for a single random choice of measurement basis $m$, considering $K$ shots, $\hat{\rho}$ can be written as~\cite{Elben2022}
\begin{equation}
    \hat{\rho}^{(m)} = \frac{1}{K} \sum_{k = 1}^K \bigotimes_{j = 1}^n \left( 3U_j^{\dagger} \big| b_j^{(m,k)} \big \rangle \! \big\langle b_j^{(m,k)} \big| U_j - \mathbb{I} \right),
    \label{eq: rho estimator}
\end{equation}
where $n$ is the number of qubits in the system, $U_j$ is the randomly selected unitary, $| b_j^{(m,k)} \rangle$ represents the post-measurement collapsed state and $\mathbb{I}$ is the identity acting on a single qubit. Following up, we turn our attention to expectation values of observables $o = \text{Tr}(O\rho)$. The prediction step empirically averages the estimator for the expected value for each measurement round $m$:
\begin{equation}
    \hat{o} = \frac{1}{M} \sum_{m = 1}^M \text{Tr} \left(O \hat{\rho}^{(m)} \right),
\end{equation}
resulting in a total number of $N = MK$ measurements. It is worth noting that the original version of classical shadows considered single-shot implementations, which is equivalent to setting $K=1$ in Eq. \eqref{eq: rho estimator}.

One of the most promising aspects of the classical shadows method is that it exhibits a sample complexity logarithmic in the number of observables. Specifically, for the estimation of $L$ observables with additive error $\varepsilon$, the total sample complexity for classical shadows is given by~\cite{Huang2020}
\begin{equation}
    N = 2\log \left(\frac{2L}{\delta} \right)  \frac{34}{\varepsilon^2} \max_{1 \leq i \leq L} \left \| O_i - \frac{\text{Tr}(O_i)}{2^n} \mathbb{I} \right \|^2_{\text{shadow}},
    \label{eq: shadow sample complexity}
\end{equation}
in which $\delta$ represents the failure probability, \textit{i.e.}, the probability that at least one of the $L$ estimated values differs from its exact counterpart by more than $\varepsilon$ in absolute terms.
Besides the exponential gain regarding the number of observables, the shadow norm in Eq. \eqref{eq: shadow sample complexity} can significantly affect the number of measurements $N$. For Pauli measurements, the sample complexity is bounded by an exponential factor in the locality $w$ of the operators, 
\begin{equation}
    N = \mathcal{O} \left( \frac{3^w \log(L)}{\varepsilon^2} \right).
    \label{eq: sample complexity bound}
\end{equation}
This represents a tradeoff between the number of observables and their locality. For instance, if one is interested in the properties of small subsystems or any sort of operators with low locality, classical shadows clearly represents a favorable approach. However, classical shadows are not appropriate for full-state tomography: if one takes $w$ to be the size of the whole system, the exponential sample complexity of standard tomography is recovered.  
%
%
\section{FALQON and Classical Shadows}
\label{sec: falqon and cs}
The FALQON (Feedback-based Algorithm for Quantum Optimization) has an association with Quantum Lyapunov Control (QLC)~\cite{magann_lyapunov_2022}. The latter was developed to map a quantum system from an arbitrary initial state to a target one by continuously adjusting the controls through a chosen feedback law~\cite{hou_optimal_2012}. In this section, we first review the original proposal of FALQON and then present our modification of this quantum algorithm. At the end of the section, we introduce the MaxCut problem, which will be used as an example of the application of the modified FALQON.  
\subsection{FALQON - A brief review}
\label{subsec: falqon review}
Nonrelativistic quantum mechanics is based on the solution of the Schrödinger equation for a given Hamiltonian and initial state~\cite{Magann2022}. 
The problem we aim to solve is defined by the total Hamiltonian $H_t = H_p + \beta(t) H_d$, in which \( H_p \) and \( H_d \) are the problem and driver Hamiltonians, respectively.  The Hamiltonian $H_p$ encodes the problem we wish to solve, whose solution is determined by its ground state, and $H_d$ drives the state of the system to explore the solution space. The time-dependent parameter \(\beta(t)\) modulates the system to drive it toward the desired target state, which minimizes the energy of the system. To achieve this, we use Quantum Lyapunov Control (QLC)~\cite{isidori_nonlinear_1995} and define an appropriate Lyapunov function  for \( \beta(t) \), guiding the system toward the desired solution. At the end of the evolution, at time $T$, we require $\beta(T) = 0$, such that the Hamiltonian $H_t$ reduces to the problem Hamiltonian $H_p$. Summarizing, the goal is to minimize the problem Hamiltonian by controlling $\beta(t)$ \cite{Magann2022}. Thus, we identify the cost function of our problem as
\begin{equation}
    C(t) \equiv \langle H_p \rangle = \langle \psi(t) | H_p | \psi(t) \rangle.
    \label{eq: cost function}
\end{equation}
To minimize the cost function, we impose that it decreases with time \cite{isidori_nonlinear_1995}: 
\begin{equation} 
\label{eq: cost evolution}
    \frac{\mathrm{d}}{\mathrm{d}t} \langle \psi(t) | H_p | \psi(t) \rangle \leq 0, \quad \forall t \geq 0. 
\end{equation}
Then, the solution to this problem is
\begin{align}
    \frac{\mathrm{d}}{\mathrm{d}t} \langle  H_p  \rangle  &=  \langle \psi(t) |i [H_t, H_p] | \psi(t) \rangle \nonumber \\
    &= A(t) \beta(t),
\end{align}
where $A(t) \equiv \langle \psi(t) | i[H_d, H_p] | \psi(t) \rangle$. In order to continuously update the parameter $\beta(t)$, we must select an appropriate feedback law. This choice allows the algorithm to remain purely quantum, avoiding issues like local minima and bypass scalability problems. The feedback control law commonly found in the literature is expressed as
\begin{equation}
    \beta(t) = -\alpha f(t, A(t)).
\end{equation}
According to Lyapunov's principle~\cite{isidori_nonlinear_1995}, the function $f$ must be continuous, with \(f(t, 0) = 0\) and satisfying the condition \(A(t) f(t, A(t)) > 0\) for all \(A(t) \neq 0\). For the simplest case, with \( \alpha = 1 \) and \( f(t, A(t)) = A(t) \), we have~\cite{magann_lyapunov_2022}
  \begin{equation}
       \beta(t) = -A(t).
       \label{eq: beta definition}
  \end{equation}
As a consequence, we ensure that $\langle H_p \rangle $ decreases monotonically, given that
\begin{equation}
    \frac{\mathrm{d}}{\mathrm{d}t}\langle H_p \rangle  = -A^2(t)  \leq 0.    
\end{equation}
To solve the proposed differential equation, we need to handle the time-dependent differential operator, which leads to a more complex problem involving the Dyson series. In addition to the complexity of solving these series, their implementation in quantum circuits is also challenging. Therefore, a viable approach is to use the  Trotter-Suzuki decomposition of the total time evolution operator~\cite{nielsen2000quantum}
\begin{equation}
    U = U_d (\beta_\ell)U_p \ldots U_d(\beta_1)U_p,
\end{equation}
in which the operators $U_p$ and $U_d$ are given by
\begin{equation} \label{eq:upud}
    \begin{aligned}
        U_p &= e^{-iH_p\Delta t}, \\
        U_d(\beta_k) &= e^{-i\beta_k H_d\Delta t},
    \end{aligned}
\end{equation}
with $\beta_k = \beta(k\tau - \Delta t)$ for $k = 1,2,...,\ell$, such that $\ell$ is the number of layers, $k$ determines the time step, and $\tau = 2\Delta t$. Notice that after each period of time $\Delta t$ the applied Hamiltonian alternates between $H_p$ and $H_d$. To ensure the validity of Eq. \eqref{eq: cost evolution}, we update the value of $\beta(t)$ as $\beta_{k+1} =- A_k$, with $A_k = \langle \psi_k | i[H_d, H_p] | \psi_k \rangle$. The parameter $\Delta t$ must be chosen sufficiently small to avoid oscillations in the cost function and $\beta (t)$ parameter, guaranteeing the convergence of the algorithm~\cite{isidori_nonlinear_1995}. The algorithm stops when the decrement in the cost function is smaller than a predefined tolerance: $C_l - C_{l+1}< \epsilon$, where $\epsilon$ is a small, positive constant.
\subsection{Classical Shadows for FALQON}
\label{subsec: mod falqon}

\begin{figure*}
    \centering
    \includegraphics[width=0.8\linewidth]{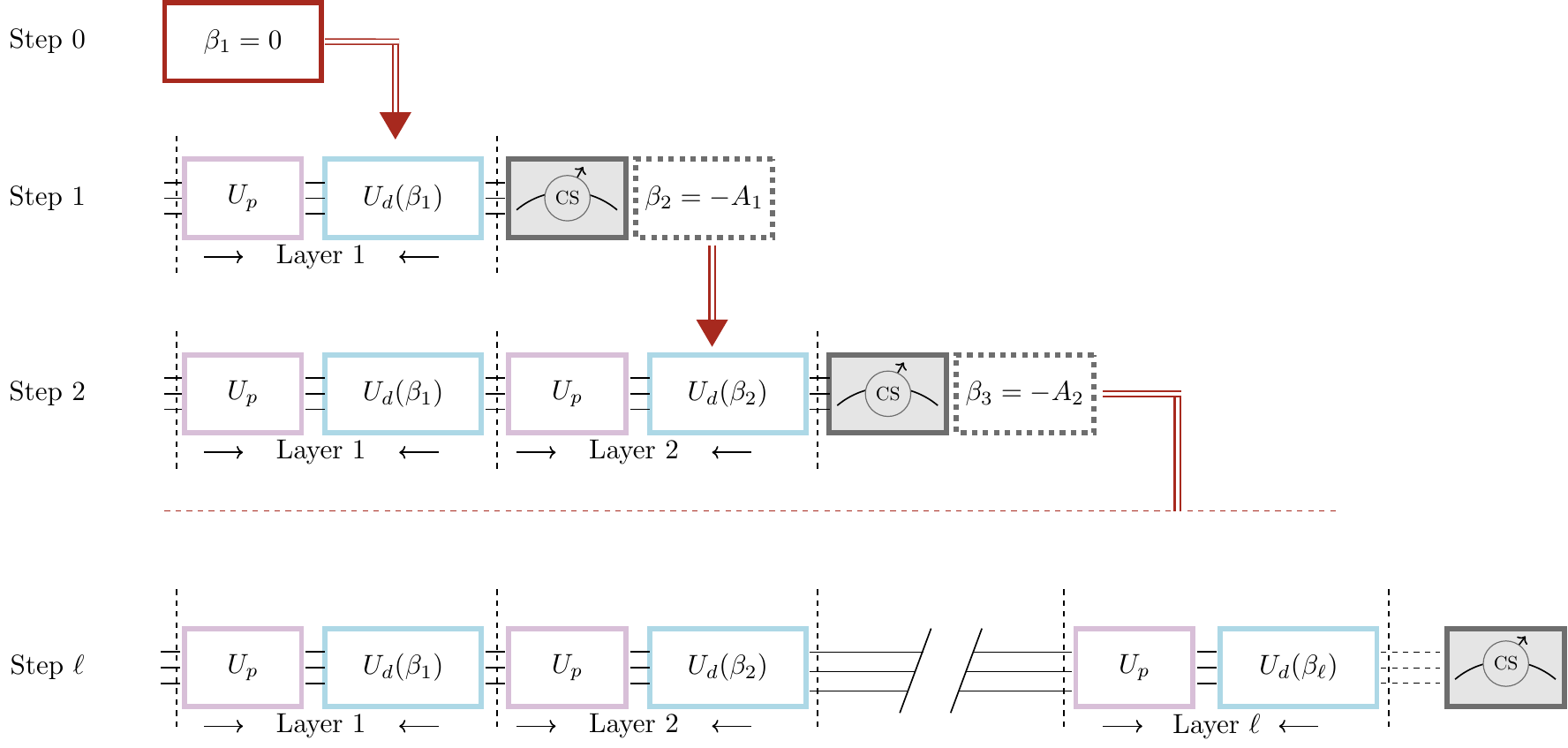} 
    \caption{Modification of the original FALQON to implement shadow measurements as a subroutine of the algorithm. The rectangles with $U_p$ and $U_d$ indicate the application of the operators defined in Eq.~\eqref{eq:upud}, while CS stands for the estimation of $A_k$ with classical shadows. The values of the control function are updated according to $\beta_{k+1} = -A_k $ at the end of each layer. Finally, after $\ell$ layers, the cost function $C_l$ is measured. If $| C_l - C_{l+1} | < \epsilon$, for a given tolerance $\epsilon$, the algorithm stops.}
    \label{fig: circuit model falqon}
\end{figure*}

FALQON relies on precise evaluations of expected values of physical observables (see Eq.~\eqref{eq: cost function} and Eq.~\eqref{eq: beta definition}), typically Pauli observables of the form $O = \bigotimes_j P_j$, where $P_j \in \{ \mathbb{I}, X, Y, Z \}$. Considering the tensor product structure of the estimators in Eq. \eqref{eq: rho estimator}, one can easily estimate the expected values of Pauli observables without the need to reconstruct the density operator of the system explicitly. We explore this property by implementing classical shadows as a subroutine for the prediction step of FALQON, as shown in Fig. \ref{fig: circuit model falqon}

In particular, this approach becomes advantageous when the operators involved in the estimation of the control function $\beta(t)$ and cost function $C(t)$ have low locality, once the shadow norm will not introduce substantial overhead in the sample complexity for estimating the observables. Implementing shadow strategies for the estimation of multiple observables has already been proposed in previous works, highlighting its versatility and usefulness in different fields \cite{Huggins2022, Blunt2024, Basheer2024, Benchen_Huang2024, Shen2024, Chan2024, Boyd2022, Ghisoni2024}. 

Ever since its first appearance, several variations and improvements have been proposed to the original protocol of classical shadows. Notably, the derandomization method~\cite{Huang2021} and locally biased procedures~\cite{Hadfield2022} are prominent examples. The first allows for efficiently estimating Pauli observables with high locality, by optimizing the choice of unitaries in the measurement procedure; the latter explores changing the probability distribution from which one samples the unitaries. 

The algorithm begins by setting \(\beta_1 = 0\), which can be achieved by a proper choice of \(|\psi_0\rangle\) as an eigenstate of $H_d$. Next, a single layer of the algorithm is implemented to prepare \(|\psi_1\rangle = U_d(\beta_1) U_p |\psi_0\rangle\), where \(U_p\) and \(U_d\) are the unitary operators described in Eq.~\eqref{eq:upud}. After that, the classical shadows protocol is used to estimate the cost function $C(t)$ and $A_1 = \langle \psi_1 | i[H_d, H_p] | \psi_1 \rangle$. This result is then used to define \(\beta_2 = -A_1\). For subsequent steps, the same procedure is repeated, iterating through the process of updating parameters and adding layers until the desired number of iterations $\ell$, is reached.  If the condition $| C_l - C_{l+1} | < \epsilon$ is satisfied, where $\epsilon$ is a predetermined tolerance, the iteration process stops, returning the approximate ground state of the system, \(|\psi_{\ell}\rangle\)  which encodes the solution to the problem. Thus, the algorithm can be executed either by checking the halting condition above or, alternatively, by predefining a sufficiently large number of iterations, $\ell$, to establish convergence. The former approach is useful when running exact simulations of the algorithm, the latter is adopted in this study, as our focus is the performance of the algorithm when measurements are taken into account.

As follows, we apply the FALQON to solve the MaxCut problem using two different strategies of measurements, the standard procedure (in which direct measurements are performed) and classical shadows.
\subsection{Application to  MaxCut Problem}
MaxCut is a classical combinatorial optimization problem that is classified as NP-hard. Given a set of $n$ vertices, the goal is to split the graph into two subsets of vertices in such a way that the number of edges connecting vertices from different subsets is maximized~\cite{mcandrew_adiabatic_2020}. The Hamiltonian associated with the MaxCut problem for an unweighted graph $\mathcal{G}$,  with $\mathcal{E}$ edges and $n$ nodes, is expressed as~\cite{Magann2022}
\begin{equation}
\label{eq: H_p and H_d}
H_p = - \frac{1}{2} \sum_{i,j \in \mathcal{E}} \left(1 - Z_i Z_j \right),  
\end{equation}
where the double sum runs over the indices $i$ and $j$ belonging to the set $\mathcal{E}$, referring to the pairs of connected vertices of the graph. Choosing the driver Hamiltonian as
\begin{equation}
H_d = \sum_{j=1}^n X_j
\end{equation}
and the initial state \(|\psi_0\rangle = |+\rangle ^{\otimes n} \), we guarantee that $\beta_1 = 0$. Then, the commutator of $H_d$ and $H_p$ reads
\begin{equation}
    \label{eq: beta commutator}
    i[{H}_d, {H}_p] = \sum_{i,j \in  \mathcal{E} } {Y}_i {Z}_j + {Z}_i {Y}_j.
\end{equation}
It is important to notice that as the MaxCut problem involves only 2-local observables ($w=2$), see Eq.~\eqref{eq: beta commutator}, the shadow norm will not introduce substantial overhead in the sample complexity for estimating the observables. 

As the cost function is determined by the expected value of the problem Hamiltonian, see Eq.~\eqref{eq: H_p and H_d}, standard measurements in the computational basis are already efficient to evaluate it. However, as classical shadows allows for the estimation of multiple observables from the same measured data, we employ its protocol in the evaluation of both cost and control functions. Regarding the control parameter $\beta_k$, we note that, as it consists of combinations of tensor products of $Y$ and $Z$, none of the operators required contains the Pauli-$X$ matrix. Then, we can bias the sampling of unitaries by simply removing the $X$-basis from the possible choices of bases. By doing so, we still have a randomized protocol, but optimized in the sense that all choices of basis will be useful for the purpose of the algorithm.

\begin{figure}[ht!]
    \centering
    \includegraphics[width = \linewidth]{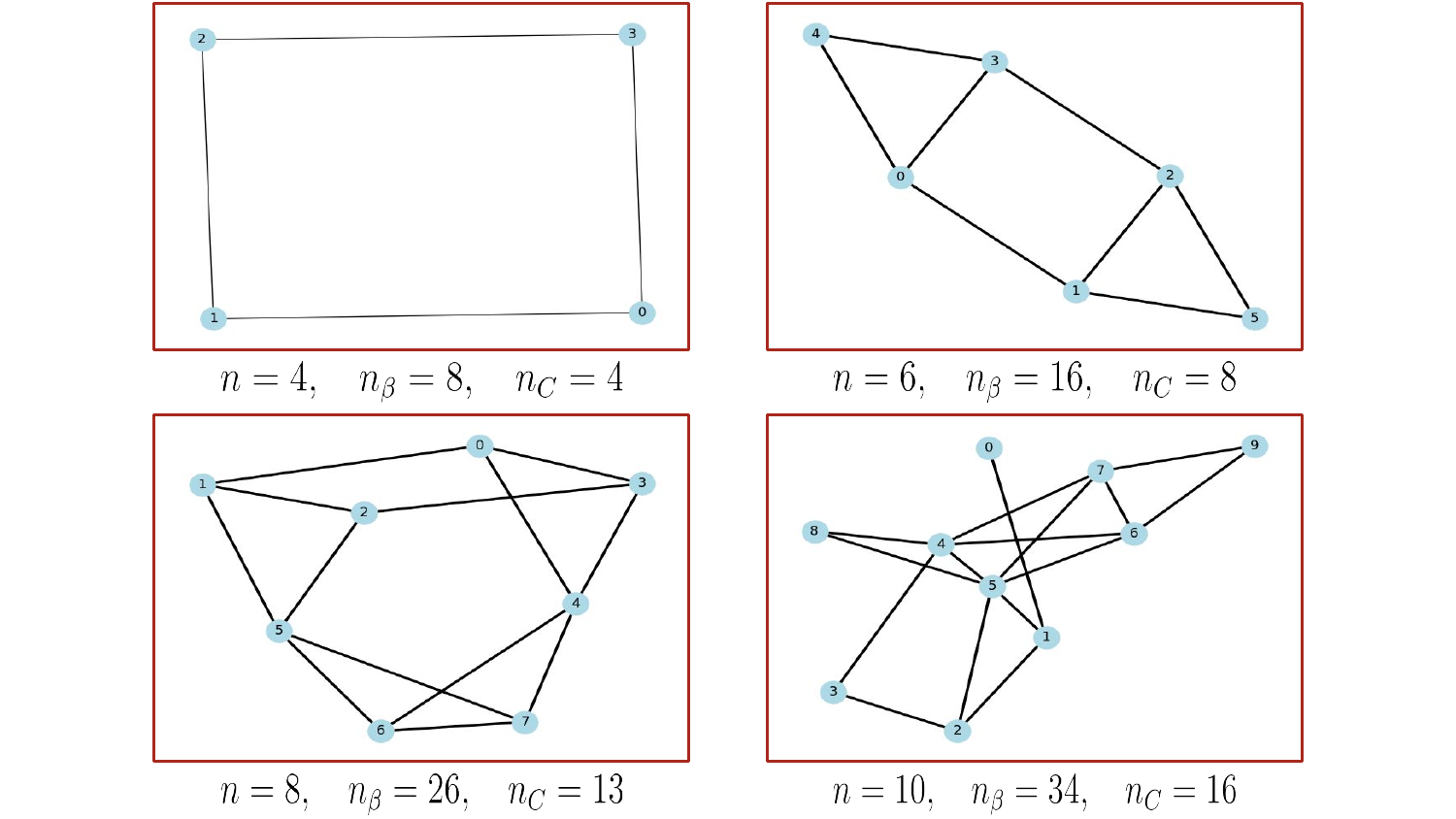} 
    \caption{Number of operators to be measured in the control function $n_{\beta}$ and cost function $n_C$ for a graph with $n \in \{4, 6, 8, 10\}$ vertices, in which $n_\beta$ and $n_C$ represent the number of operators needed to estimate the parameter $\beta$ and the cost function $C$, respectively, considering different geometries of the MaxCut problem.} 
    \label{fig: graph_operators}
\end{figure}

Below we analyzed the MaxCut problem for a few selected geometries, with $n \in \{4, 6, 8, 10\}$ vertices, as illustrated in Fig.~\ref{fig: graph_operators}, along with the required number of operators to evaluate the cost function ($n_C$) and the control parameter ($n_\beta$), according to Eq.~\eqref{eq: H_p and H_d} and Eq.~\eqref{eq: beta commutator}, respectively. We started with a simple rectangular geometry and progressively increased the number of vertices and inner edges to test the algorithm on graphs with low symmetry. The results are presented in the next section, along with an analysis of the scaling law for the number of measurements using classical shadows to solve the MaxCut problem for complete graphs.
%
\section{Results and discussion}
\label{sec: results}
The four cases of the MaxCut problem shown in Fig.~\ref{fig: graph_operators} are analyzed in this work. We implement the FALQON algorithm using two measurement procedures, direct and shadow estimations, through Ket quantum programming platform~\cite{ket2021}, and compare them to the exact calculations. The first represents the case of non-optimized measurements, which we refer to as direct measurements. In this case, we take enough shots to obtain sufficiently accurate estimations of each observable individually; the second uses the previously described biased classical shadows, allowing for the reuse of information from the randomized measurements in the classical post-processing. For clarification purposes, by exact value of an observable $O$ we mean that its expected value is evaluated numerically using the expression $\langle \psi_k | O |\psi_k \rangle$, in which $|\psi_k \rangle$ is the output state of the quantum circuit after $k$ layers. 

\begin{figure*}[tbp]
    \centering
    \begin{minipage}{.49\textwidth} 
        \centering
        \resizebox{\linewidth}{!}{\includegraphics{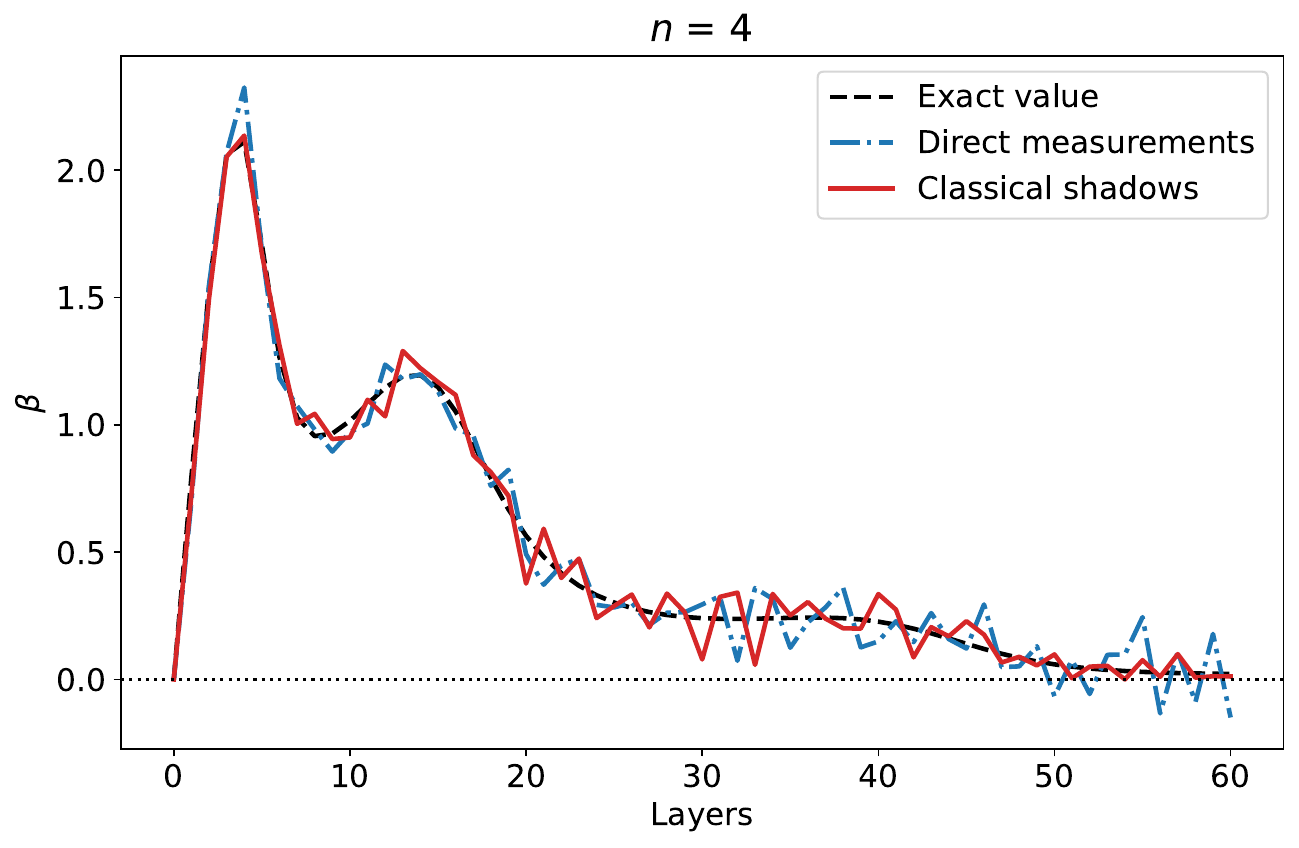}}
    \end{minipage} 
    \begin{minipage}{.49\textwidth} 
        \centering
        \resizebox{\linewidth}{!}{\includegraphics{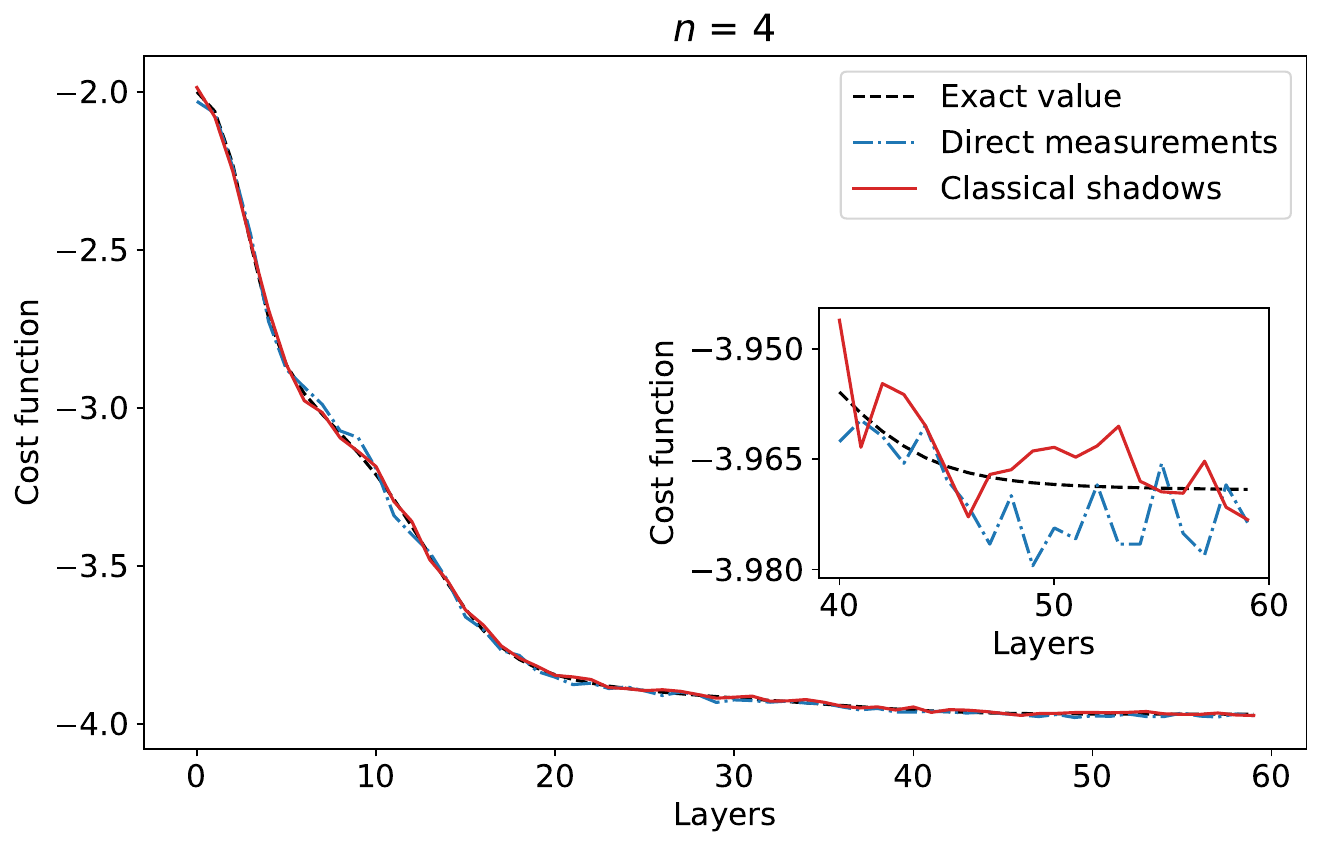}}
    \end{minipage}
    
    \begin{minipage}{.49\textwidth} 
        \centering
        \resizebox{\linewidth}{!}{\includegraphics{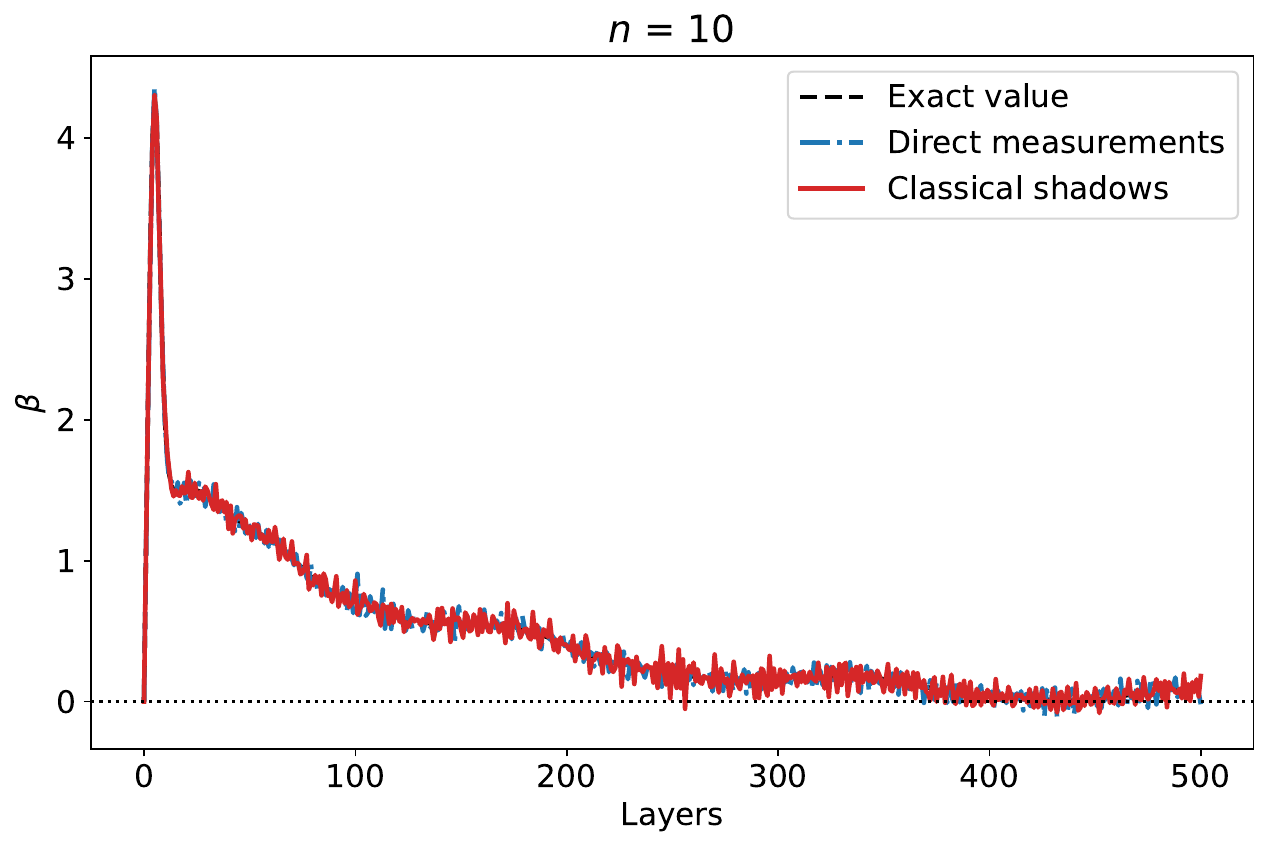}}
    \end{minipage}
    \begin{minipage}{.49\textwidth} 
        \centering
        \resizebox{\linewidth}{!}{\includegraphics{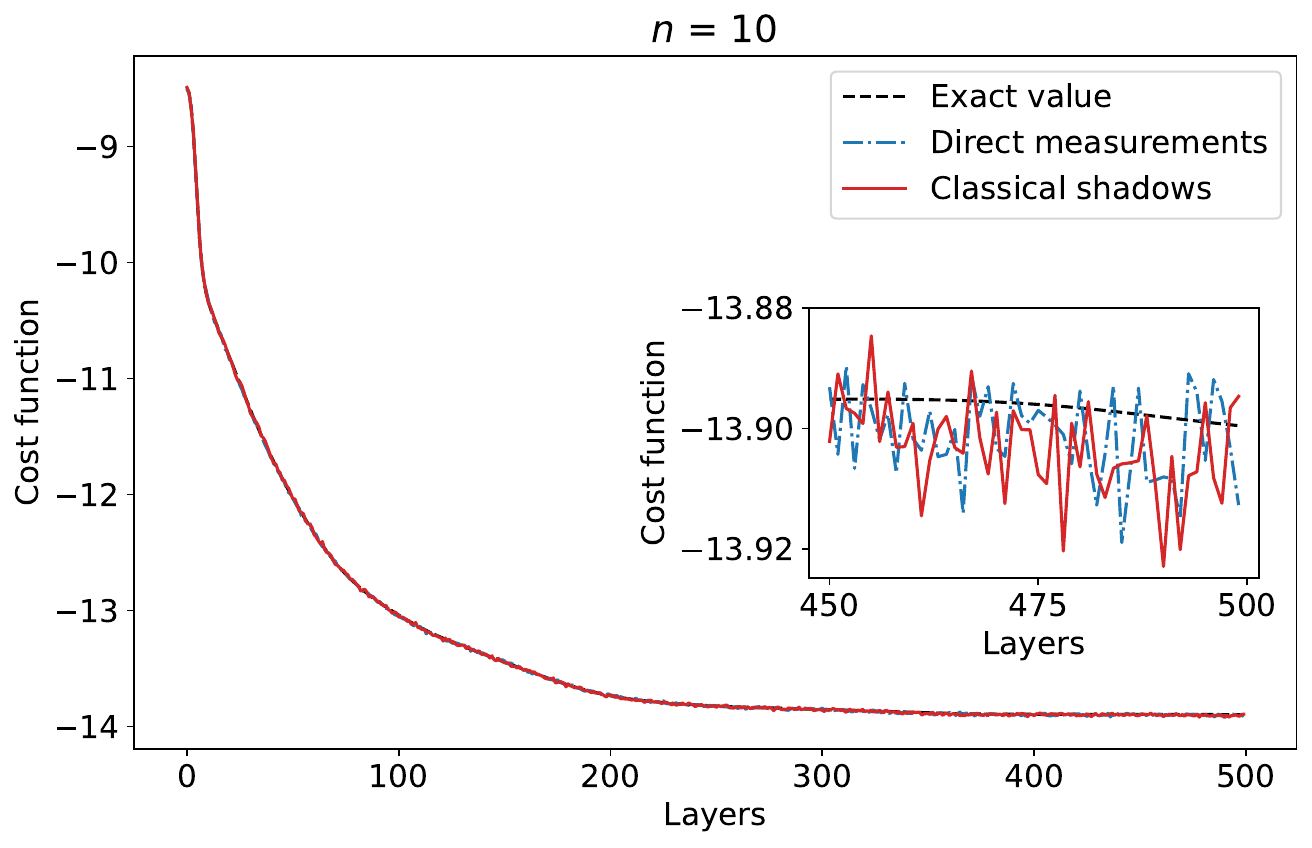}} 
    \end{minipage}
    \caption{Behavior of the control function $\beta (t)$ and the cost function $C(t)$ for the graphs described in Fig. \ref{fig: graph_operators} with $n=4$ and $n=10$ vertices. The estimated values of $\beta (t)$ and $C(t)$ are obtained through different strategies (direct measurements and classical shadows) and compared to the exact values for different numbers of layers. The number of measurements necessary to obtain these results was determined by imposing a maximum mean additive error of $0.01$ between the estimated and exact cost functions. The insets in the cost function figures highlight the oscillations caused by the measurement process.}
    \label{fig: results mean error}
\end{figure*}

To evaluate the performance of the proposed measurement procedure, we establish a criterion based on the absolute difference between the estimated and exact values for the cost function, denoted by $\Delta C$. Considering all the layers used to run the algorithm, we require the average of $\Delta C$ not to exceed a specified constant $\texttt{err}$, chosen to be sufficiently small to guarantee the algorithm's convergence:
\begin{equation}
    \overline{\Delta C} = \frac{1}{\ell} \sum_{i=1}^{\ell} \Delta C_i  \leq \texttt{err}.
    \label{eq: mean error}
\end{equation}
This criterion is used to account for the oscillations in the values of the cost function, given the probabilistic nature of the outputs of quantum measurements. We set $\texttt{err} = 0.01$ and simulate the algorithm in Ket quantum programming platform, doubling the total number of measurements until we reach $\overline{\Delta C} \leq \texttt{err}$, to ensure the appropriate convergence for both measurement procedures. For shadow estimation, we fixed the number of repetitions to $K = 128$, increasing the number of random unitaries $M$; for direct estimation, the total number of measurements is simply the number of shots. The evolution of the control parameter $\beta$ and the cost function $C$, plotted as a function of the number of layers, is illustrated in Fig. \ref{fig: results mean error} for the problems with $n=4$ and $n=10$ nodes, respectively. We observe that the values of the control and cost functions are randomly distributed around the exact value (black dashed line) when the direct measurements (blue dotted and dashed line) and classical shadows (red continuous line) methods are employed. This arises from the fundamental impossibility of predicting the exact outcome of a measurement in a quantum system. Also, for smaller graphs the solution is easily obtained, which is reflected in the faster convergence of the cost function, requiring a smaller number of layers.
Now, we focus our attention on the total number of measurements required to satisfy the criterion given by Eq. (\ref{eq: mean error})  with $\texttt{err} = 0.01$, displaying the results in Table \ref{table: number of measurements}. In cases with a smaller number of nodes, such as $n=4$, the same number of measurements sufficed for both methods, suggesting that, for problems with fewer vertices, the efficiency of both methods is equivalent. However, as the number of nodes increases, we can observe a clear difference between the two approaches. For $n=6$, the direct measurements approach required $8$ times the number of measurements needed with biased classical shadows. Such a difference becomes even more pronounced for $n=8$ and $n=10$, when the direct estimation of observables required surpasses one million shots, $16$ times the required measurements needed with classical shadows. The considerable difference in the measurement budget presented in Table \ref{table: number of measurements} suggests that the shadow approach is particularly well-suited for large-scale optimization problems that rely on the estimation of many expected values of low locality operators.
\begin{table}
  \centering
  \caption{Number of measurements per layer, ensuring ${\texttt{err} \leq 0.01}$ for the average cost function $\overline{\Delta C}$ (see Eq.~\eqref{eq: mean error}).}
   \begin{tabular}{r @{\hspace{0.75cm}} r @{\hspace{0.75cm}} r}
    \toprule
    \textbf{\makecell{Number\\of Nodes}} & \textbf{\makecell{Classical\\Shadows}} & \textbf{\makecell{Direct\\Measurements}} \\
    \midrule
    4  & 16,384   & 16,384     \\
    6  & 32,768   & 262,144    \\
    8  & 32,768   & 524,288    \\
    10 & 65,536   & 1,048,576  \\
    \bottomrule
  \end{tabular}
  \label{table: number of measurements}
\end{table}
We recall that the cost function is the expected value of the problem Hamiltonian, $H_p$, whose evolution relies on accurate estimations for the $\beta$ parameter. Thus, by fixing an average mean error for the cost function, we are also implicitly requiring precise estimates of $\beta$. Moreover, for both $\beta$ and the cost function, we are dealing with expected values of multiple 2-local observables, which guarantees that the sample complexity presented in Section~\ref{sec: shadows} will not grow exponentially. The better performance of classical shadows can be understood by its ability to extract information during the post-processing of data, once the measurement outcome of a single Pauli string on $n$ qubits can be used to estimate several different 2-local observables. In other words, it allows for the reconstruction of multiple observables from a fixed set of measurements. 

Our numerical analysis points to an increasing difference in performance between the two methods as the number of nodes grows. Given that, for each new edge on a graph, two more operators are needed to estimate $\beta$ and one more to estimate the cost function $C$, this difference in the measurement budget tends only to increase, as more connected graphs are considered. Therefore, the classical shadows protocol stands as a promising tool to tackle scalability challenges in quantum optimization.

We extend our study by imposing a stricter criterion on observable estimation, now focusing on the control parameter $\beta$ and analyzing complete (fully connected) graphs. The choice of complete graphs is motivated by their higher number of edges, which increases the number of observables to be estimated in the control parameter $\beta$. The total number of measurements required to achieve accurate results with FALQON poses an important role when seeking quantum advantage. For graphs with many vertices, a large number of layers may be required, making such number a critical factor. In this context, characterizing the scaling law of the number of measurements per layer, as a function of the number of operators, can help extrapolate results to larger MaxCut instances. 

\begin{figure}
    \centering    
    \includegraphics[width = 0.9\linewidth]{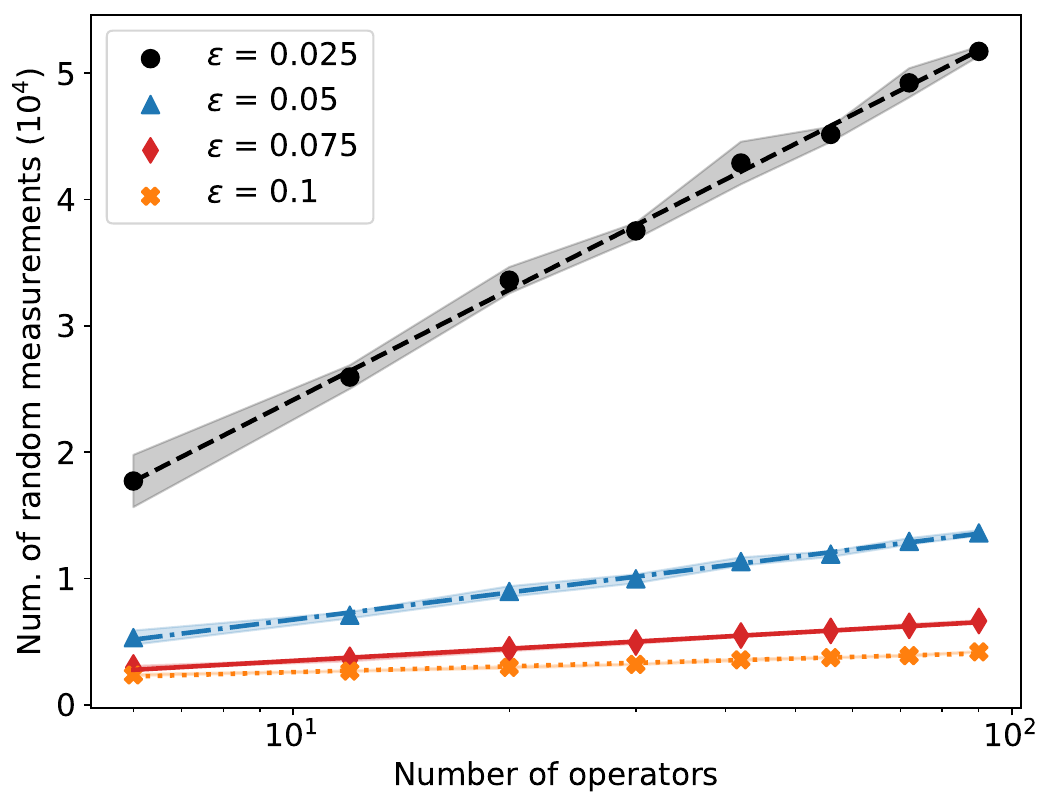}
    \caption{Average number of measurements per layer as a function of the number of operators in the control function to solve the MaxCut problem with shadow measurements for different additive errors in the estimation of the observables. The shaded regions are the standard deviations of the required number of measurements, over ten independent runs.}
    \label{fig: scale law}
\end{figure}

The sample complexity for estimating a set of observables with classical shadows is upper bounded by Eq.~\eqref{eq: sample complexity bound}, where $\varepsilon$ is the additive error. Since for this particular problem we excluded the $X$-basis from the set of possible choices, the $3^w$ factor reduces to $2^w$. This reduction can be understood in probabilistic terms, given that each qubit now has only two possible basis choices \cite{Ippoliti2024}. Consequently, the number of measurements to estimate $L$ operators is expected to scale as $N = \mathcal{O} \left( 4 \log(L) / \varepsilon^2 \right)$, as we are dealing with 2-local operators. We fixed four values of $\varepsilon$ and required all expected values in the estimation of $\beta$ to fall within precision $\varepsilon$. If this condition was not met, the number of measurements was increased. We adjust our results by applying a logarithmic fit, 
\begin{equation}
N = A \frac{4 \log_{10}(L)}{\varepsilon^2} + B,
\end{equation}
where $N$ represents the mean number of measurements per layer. The findings related to the scaling law are depicted in Fig. \ref{fig: scale law}, and the fit parameters are listed in Table \ref{table: log fit}. Because we seek an upper bound on the scaling law, we consider the maximum value of the parameter $A$ and take $\lceil A \rceil = 5$, resulting in the same factor for all the cases studied. This leads to a scaling law of the form $20\log_{10}(L)/\varepsilon^2$, up to an additive constant.
\begin{table}
\centering
\caption{Logarithmic fit parameters for the curves in Fig. \ref{fig: scale law}. The parameter $A$ exhibits the scaling factor in the sample complexity for the solution to MaxCut with FALQON.}
\begin{tabular}{c @{\hspace{0.75cm}} c @{\hspace{0.75cm}} r}
\toprule
$\bm{\epsilon}$ & $\bm{A}$  & $\bm{B}$ \\
\midrule
$ 0.025$ & $4.533$ & $-4886$ \\
$ 0.050$ & $4.458$ & $-385$ \\
$ 0.075$ & $4.500$ & $286$  \\
$ 0.100$ & $3.866$ & $1043$  \\
\bottomrule
\end{tabular}
\label{table: log fit}
\end{table}
\section{Conclusion}
\label{sec: conclusions}
We studied the performance of FALQON applied to the solution of the MaxCut problem for different graph geometries in a more realistic scenario. Two methods to estimate the observables involved in the feedback law were used: direct measurements, which is a brute-force approach to measure all the given operators, and a biased version of classical shadows, obtained by simply removing the Pauli $X$ basis from the bases ensemble. To evaluate the performance of these approaches, we fixed a threshold $\texttt{err}$, for the absolute mean difference per layer, between the estimated values and the exact ones. The numerical results for $\texttt{err} = 0.01$ showed a significant advantage when using classical shadows compared to the direct measurement procedure. This advantage becomes clear for highly connected graphs or graphs with a high number of vertices, leading to a large number of operators to be measured. Specifically, for $n = 8$ and $n = 10$, shadow estimation requires 16 times fewer measurements than the direct method. Also, in the case of complete graphs, the logarithmic scaling of the number of measurements to estimate the expected values of all Pauli strings in the control function was analyzed. This inspection allowed us to obtain the constants that determine the scale law for solving MaxCut with FALQON, which can be used to extrapolate experimental budget estimations for bigger problems. These results point to possible applications of classical shadows as subroutines in quantum algorithms to estimate Pauli strings with low locality. In particular, any algorithm that requires the estimation of many expected values with low locality can benefit from randomized classical shadows.
\begin{acknowledgments}
The authors gratefully acknowledge the financial support provided by the Brazilian funding agencies Conselho Nacional de Desenvolvimento Científico e Tecnológico (CNPq) and Coordenação de Aperfeiçoamento de Pessoal de Nível Superior (CAPES). EID, LB, and JPE also acknowledge the support of the Instituto Nacional de Ciência e Tecnologia de Informação Quântica (INCT-IQ). Additionally, EID and LB thank CNPq for funding through Grant No. 408341/2022-0, and EID acknowledges further support from CNPq under Grant No. 409673/2022-6.
\end{acknowledgments}
\bibliography{apssamp}
\end{document}